\documentclass[aps,prl,preprint,superscriptaddress]{revtex4-1}

\usepackage{graphicx}

\usepackage{amssymb,amssymb}

\usepackage{lineno}

\usepackage{url}
\usepackage{ifthen}
\usepackage{subfigure}

\def\mttbar {t\, \overline{t}}
\def\ttbar {$\mttbar$}
\def\GeVc  {$\rm GeV\!/{\it c}$}

\def\GeVccm {\rm GeV\!/{\it c}^2}
\def\GeVcc {$\GeVccm$}
\def\mMEt{\not\kern-.35em {E_T}}
\def\MEt{\hbox{$\mMEt$}}
\def\invfb {{fb$^{-1}$}}

\newcommand{\pT}{p_{T}}

\newcommand{\ET}{E_{_T}}

\newcommand{\met}    {\mbox{$\protect \raisebox{.3ex}{$\not$}\ET$}}

\newcommand{\mSigmet}{S_{MET}}

\newcommand{\antikt}{{\it anti-k$_t$}}

\def\mjet#1{m^{jet#1}}
\newcommand{\sampleLum}{5.95~\invfb}
\newcommand{\Nvtx}{N_{vtx}}
\newcommand{\angularity}{\tau_{-2}}
\newcommand{\planarflow}{Pf}
\def\Pythia{{\sc pythia}}
\def\Fastjet{{\sc fastjet}}

\begin{document}
\preprint{FERMILAB-PUB-11-297-E-PPD}

\title{Study of Substructure of High Transverse Momentum Jets Produced in Proton-Antiproton 
Collisions at $\sqrt{s}=1.96$\ TeV}


\affiliation{Institute of Physics, Academia Sinica, Taipei, Taiwan 11529, Republic of China} 
\affiliation{Argonne National Laboratory, Argonne, Illinois 60439, USA} 
\affiliation{University of Athens, 157 71 Athens, Greece} 
\affiliation{Institut de Fisica d'Altes Energies, ICREA, Universitat Autonoma de Barcelona, E-08193, Bellaterra (Barcelona), Spain} 
\affiliation{Baylor University, Waco, Texas 76798, USA} 
\affiliation{Istituto Nazionale di Fisica Nucleare Bologna, $^{aa}$University of Bologna, I-40127 Bologna, Italy} 
\affiliation{University of California, Davis, Davis, California 95616, USA} 
\affiliation{University of California, Los Angeles, Los Angeles, California 90024, USA} 
\affiliation{Instituto de Fisica de Cantabria, CSIC-University of Cantabria, 39005 Santander, Spain} 
\affiliation{Carnegie Mellon University, Pittsburgh, Pennsylvania 15213, USA} 
\affiliation{Enrico Fermi Institute, University of Chicago, Chicago, Illinois 60637, USA}
\affiliation{Comenius University, 842 48 Bratislava, Slovakia; Institute of Experimental Physics, 040 01 Kosice, Slovakia} 
\affiliation{Joint Institute for Nuclear Research, RU-141980 Dubna, Russia} 
\affiliation{Duke University, Durham, North Carolina 27708, USA} 
\affiliation{Fermi National Accelerator Laboratory, Batavia, Illinois 60510, USA} 
\affiliation{University of Florida, Gainesville, Florida 32611, USA} 
\affiliation{Laboratori Nazionali di Frascati, Istituto Nazionale di Fisica Nucleare, I-00044 Frascati, Italy} 
\affiliation{University of Geneva, CH-1211 Geneva 4, Switzerland} 
\affiliation{Glasgow University, Glasgow G12 8QQ, United Kingdom} 
\affiliation{Harvard University, Cambridge, Massachusetts 02138, USA} 
\affiliation{Division of High Energy Physics, Department of Physics, University of Helsinki and Helsinki Institute of Physics, FIN-00014, Helsinki, Finland} 
\affiliation{University of Illinois, Urbana, Illinois 61801, USA} 
\affiliation{The Johns Hopkins University, Baltimore, Maryland 21218, USA} 
\affiliation{Institut f\"{u}r Experimentelle Kernphysik, Karlsruhe Institute of Technology, D-76131 Karlsruhe, Germany} 
\affiliation{Center for High Energy Physics: Kyungpook National University, Daegu 702-701, Korea; Seoul National University, Seoul 151-742, Korea; Sungkyunkwan University, Suwon 440-746, Korea; Korea Institute of Science and Technology Information, Daejeon 305-806, Korea; Chonnam National University, Gwangju 500-757, Korea; Chonbuk National University, Jeonju 561-756, Korea} 
\affiliation{Ernest Orlando Lawrence Berkeley National Laboratory, Berkeley, California 94720, USA} 
\affiliation{University of Liverpool, Liverpool L69 7ZE, United Kingdom} 
\affiliation{University College London, London WC1E 6BT, United Kingdom} 
\affiliation{Centro de Investigaciones Energeticas Medioambientales y Tecnologicas, E-28040 Madrid, Spain} 
\affiliation{Massachusetts Institute of Technology, Cambridge, Massachusetts 02139, USA} 
\affiliation{Institute of Particle Physics: McGill University, Montr\'{e}al, Qu\'{e}bec, Canada H3A~2T8; Simon Fraser University, Burnaby, British Columbia, Canada V5A~1S6; University of Toronto, Toronto, Ontario, Canada M5S~1A7; and TRIUMF, Vancouver, British Columbia, Canada V6T~2A3} 
\affiliation{University of Michigan, Ann Arbor, Michigan 48109, USA} 
\affiliation{Michigan State University, East Lansing, Michigan 48824, USA}
\affiliation{Institution for Theoretical and Experimental Physics, ITEP, Moscow 117259, Russia}
\affiliation{University of New Mexico, Albuquerque, New Mexico 87131, USA} 
\affiliation{Northwestern University, Evanston, Illinois 60208, USA} 
\affiliation{The Ohio State University, Columbus, Ohio 43210, USA} 
\affiliation{Okayama University, Okayama 700-8530, Japan} 
\affiliation{Osaka City University, Osaka 588, Japan} 
\affiliation{University of Oxford, Oxford OX1 3RH, United Kingdom} 
\affiliation{Istituto Nazionale di Fisica Nucleare, Sezione di Padova-Trento, $^{bb}$University of Padova, I-35131 Padova, Italy} 
\affiliation{LPNHE, Universite Pierre et Marie Curie/IN2P3-CNRS, UMR7585, Paris, F-75252 France} 
\affiliation{University of Pennsylvania, Philadelphia, Pennsylvania 19104, USA}
\affiliation{Istituto Nazionale di Fisica Nucleare Pisa, $^{cc}$University of Pisa, $^{dd}$University of Siena and $^{ee}$Scuola Normale Superiore, I-56127 Pisa, Italy} 
\affiliation{University of Pittsburgh, Pittsburgh, Pennsylvania 15260, USA} 
\affiliation{Purdue University, West Lafayette, Indiana 47907, USA} 
\affiliation{University of Rochester, Rochester, New York 14627, USA} 
\affiliation{The Rockefeller University, New York, New York 10065, USA} 
\affiliation{Istituto Nazionale di Fisica Nucleare, Sezione di Roma 1, $^{ff}$Sapienza Universit\`{a} di Roma, I-00185 Roma, Italy} 

\affiliation{Rutgers University, Piscataway, New Jersey 08855, USA} 
\affiliation{Texas A\&M University, College Station, Texas 77843, USA} 
\affiliation{Istituto Nazionale di Fisica Nucleare Trieste/Udine, I-34100 Trieste, $^{gg}$University of Udine, I-33100 Udine, Italy} 
\affiliation{University of Tsukuba, Tsukuba, Ibaraki 305, Japan} 
\affiliation{Tufts University, Medford, Massachusetts 02155, USA} 
\affiliation{University of Virginia, Charlottesville, Virginia 22906, USA}
\affiliation{Waseda University, Tokyo 169, Japan} 
\affiliation{Wayne State University, Detroit, Michigan 48201, USA} 
\affiliation{University of Wisconsin, Madison, Wisconsin 53706, USA} 
\affiliation{Yale University, New Haven, Connecticut 06520, USA} 
\author{T.~Aaltonen}
\affiliation{Division of High Energy Physics, Department of Physics, University of Helsinki and Helsinki Institute of Physics, FIN-00014, Helsinki, Finland}
\author{R.~Alon$^z$}
\affiliation{Institute of Particle Physics: McGill University, Montr\'{e}al, Qu\'{e}bec, Canada H3A~2T8; Simon Fraser University, Burnaby, British Columbia, Canada V5A~1S6; University of Toronto, Toronto, Ontario, Canada M5S~1A7; and TRIUMF, Vancouver, British Columbia, Canada V6T~2A3} 
\author{B.~\'{A}lvarez~Gonz\'{a}lez$^w$}
\affiliation{Instituto de Fisica de Cantabria, CSIC-University of Cantabria, 39005 Santander, Spain}
\author{S.~Amerio}
\affiliation{Istituto Nazionale di Fisica Nucleare, Sezione di Padova-Trento, $^{bb}$University of Padova, I-35131 Padova, Italy} 

\author{D.~Amidei}
\affiliation{University of Michigan, Ann Arbor, Michigan 48109, USA}
\author{A.~Anastassov}
\affiliation{Northwestern University, Evanston, Illinois 60208, USA}
\author{A.~Annovi}
\affiliation{Laboratori Nazionali di Frascati, Istituto Nazionale di Fisica Nucleare, I-00044 Frascati, Italy}
\author{J.~Antos}
\affiliation{Comenius University, 842 48 Bratislava, Slovakia; Institute of Experimental Physics, 040 01 Kosice, Slovakia}
\author{G.~Apollinari}
\affiliation{Fermi National Accelerator Laboratory, Batavia, Illinois 60510, USA}
\author{J.A.~Appel}
\affiliation{Fermi National Accelerator Laboratory, Batavia, Illinois 60510, USA}
% ...
\collaboration{CDF Collaboration}
\thanks{With visitors from $^a$Istituto Nazionale di Fisica Nucleare, Sezione di Cagliari, 09042 Monserrato (Cagliari), Italy,
$^b$University of CA Irvine, Irvine, CA  92697, USA,
$^c$University of CA Santa Barbara, Santa Barbara, CA 93106, USA,
$^d$University of CA Santa Cruz, Santa Cruz, CA  95064, USA,
$^e$CERN,CH-1211 Geneva, Switzerland,
$^f$Cornell University, Ithaca, NY  14853, USA, 
$^g$University of Cyprus, Nicosia CY-1678, Cyprus, 
$^h$Office of Science, U.S. Department of Energy, Washington, DC 20585, USA,
$^i$University College Dublin, Dublin 4, Ireland,
$^j$University of Fukui, Fukui City, Fukui Prefecture, Japan 910-0017,
$^k$Universidad Iberoamericana, Mexico D.F., Mexico,
$^l$Iowa State University, Ames, IA  50011, USA,
$^m$University of Iowa, Iowa City, IA  52242, USA,
$^n$Kinki University, Higashi-Osaka City, Japan 577-8502,
$^o$Kansas State University, Manhattan, KS 66506, USA,
$^p$University of Manchester, Manchester M13 9PL, United Kingdom,
$^q$Queen Mary, University of London, London, E1 4NS, United Kingdom,
$^r$University of Melbourne, Victoria 3010, Australia,
$^s$Muons, Inc., Batavia, IL 60510, USA,
$^t$Nagasaki Institute of Applied Science, Nagasaki, Japan, 
$^u$National Research Nuclear University, Moscow, Russia,
$^v$University of Notre Dame, Notre Dame, IN 46556, USA,
$^w$Universidad de Oviedo, E-33007 Oviedo, Spain, 
$^x$Texas Tech University, Lubbock, TX  79609, USA,
$^y$Universidad Tecnica Federico Santa Maria, 110v Valparaiso, Chile,
$^z$Weizmann Institute of Science, Rehovot, Israel,
$^{aa}$Yarmouk University, Irbid 211-63, Jordan,
$^{hh}$On leave from J.~Stefan Institute, Ljubljana, Slovenia, 
}
\noaffiliation



\date{\today}


\begin{abstract}
A study of the substructure of jets with transverse momentum greater than 400~\GeVc\ produced in proton-antiproton collisions at a center-of-mass energy of 1.96 TeV at the Fermilab Tevatron Collider and recorded by the CDF II detector is presented. 
The distributions of the jet mass, angularity, and planar flow are measured for the first time in a sample with an integrated luminosity of \sampleLum. 
The observed substructure for high mass jets is consistent with predictions from perturbative quantum chromodynamics.
  \end{abstract}

%
%

\maketitle
%
%
The 
study of high transverse momentum ($p_T$) massive jets produced in proton-antiproton ($p\bar{p}$) interactions 
 provides an important test of perturbative QCD (pQCD) and gives insight into the parton showering mechanism
(see e.g.,
\cite{Ellis:2007ib,
Salam:2009jx}\
for recent reviews). 
Furthermore, massive boosted jets constitute an important background in searches for various new physics models~\cite{Butterworth:2002tt,
Agashe:2006hk,
Lillie:2007yh,
Butterworth:2009qa},
the Higgs boson~\cite{Butterworth:2008iy}, 
and highly boosted top quark production.
Particularly relevant is the case where the decay of a heavy resonance produces high-$p_T$\ top quarks that decay hadronically.  
In all these cases, the hadronic decay products can be detected as a single jet with 
a large mass and internal substructure that differs on average from pQCD jets once the jet $\pT$\ is
greater than 400-500~\GeVc.
However, experimental studies of the substructure of high $\pT$\ jets at the Tevatron have been limited to jets with $\pT<400$~\GeVc~\cite{D0:2002SubClusterPRD,CDF:2005JetEnergyFlowPRD}; recently results with higher $\pT$\ jets produced at the Large Hadron Collider have been published \cite{Aad:2011kq}.
%

Jets produced through QCD processes with large mass are expected to arise
predominantly through a process of single hard gluon emission from a high $\pT$\ quark or gluon \cite{Almeida:2008tp}.  
The probability of this process is given by the jet function, $J (\mjet{},p_T,R)$,
for which a simple next-to-leading-order (NLO) approximation is
\begin{equation}\label{Jth}
J (\mjet{},p_T,R) \simeq \alpha_s (p_T)  \frac{  4\,  C_{q,g}}{\pi \mjet{}} \log\left(\frac{R \cdot p_T}{\mjet{}} \right)\,,\end{equation}
where $\mjet{}$\ is the jet mass,  $\alpha_s(p_T) $ is the strong coupling, $C_{q,g}=4/3$\ and 3 for quark and gluon jets, respectively, and $R$\ is the cone radius used to 
define the jet~\cite{Almeida:2008tp}. 
The approximation holds for $m^{jet} \ll R \cdot p_{T}$.
Although uncertainties from  higher-order corrections are $\sim 30$\%, it predicts both the shape of the spectrum and the fraction of jets with masses greater than about 100~\GeVcc.
Two other jet substructure variables insensitive to  soft radiation at high jet mass 
are angularity and planar flow~\cite{Berger:2003iw,Almeida:2008yp,Ellis:2010rw,Thaler:2008ju,pythia}.
The angularity is defined as 
\begin{equation}
\angularity(R,p_T) \equiv \frac{1}{\mjet{}} \sum_{i \in jet} E_i\, \sin^{-2} \theta_i\,
\left[ \,1 - \cos \theta_i\, \right]^{3}
\, ,
\label{tauadef}
\end{equation}
where the sum is over the constituents in the jet cluster, $E_i$\ is the energy and $\theta_i$\ is the angle of each constituent relative to the jet axis.
It is sensitive to radiation near the edge of the cone and has a characteristic shape for QCD jets.  
Planar flow is defined as 
\begin{equation}
\planarflow\equiv 
\frac{4 \lambda_1 \lambda_2}{(\lambda_1 + \lambda_2)^2} ,
\end{equation}
where $\lambda_{1,2}$ are the eigenvalues of the two-dimensional moment matrix
\begin{equation}
I^{kl}_{w}=\frac {1}{\mjet{}} \sum_{i \in jet} E_i \frac{p_{i,k}}{E_i}\,\frac{p_{i,l}}{E_i}\, ,
\end{equation}
in which $p_{i,k}$ is the $k^{th}$ component of the jet constituent's transverse energy relative to the 
jet  axis, {\it i.e.} in one of the two directions that span the plane perpendicular to the jet direction.
Jets with three of more energetic constituents, such as those arising from a boosted top quark, are more planar with $\planarflow \sim1$, compared with massive QCD jets where the energy flow is along the line defined by the two final-state partons and $\planarflow\sim0$.
Both of these variables are perturbatively calculable.

We report in this Letter the first measurement of the jet mass distribution for jets with $\pT>400$~\GeVc\ produced in 1.96~TeV $p\bar{p}$\ collisions at the Fermilab Tevatron Collider and recorded by the CDF II detector.
We also measure for jets with masses greater than $90$~\GeVcc\ their angularity and planar flow distributions.
We use the Midpoint cone algorithm \cite{Blazey:2000midpoint}\ to reconstruct jets using the \Fastjet\ program \cite{Cacciari:2006}\ and the \antikt\ algorithm \cite{antikT}, allowing for a direct
comparison of cone and recombination algorithms.

The CDF II detector  \cite{CDF:2005Jpsi}\ consists of a solenoidal charged particle spectrometer surrounded by
a calorimeter and muon system.  
Charged particle momenta are measured over  $|\eta| < 1.1$.
The calorimeter covers the region $|\eta| < 3.6$, with the region $|\eta|<1.1$\  segmented into towers of size $\Delta\eta\times\Delta\phi=0.11\times0.26$~\footnote{We use a coordinate system where $\phi$\ and $\theta$\ are the azimuthal and polar angles around the proton beam axis.  
The pseudorapidity is $\eta = -\ln\tan(\theta/2)$\ and 
$R=\sqrt{(\delta\eta)^2 + (\delta\phi)^2}$.  
}. 
The calorimeter system is used to measure jets and missing transverse energy (\MEt) defined as
\begin{equation}
\not\!\! \vec{E}_T = - \sum_{i} E_T^i \hat{n}_i,
\end{equation}
where the sum is over the calorimeter towers with $|\eta| < 3.6$\ and 
$\hat{n}_i$ is a unit vector perpendicular to the beam axis and pointing at the i$^{th}$ calorimeter tower. 
We also define $\mMEt =|{\not\!\!\vec{E}_T}|$.
The 4-momentum of a jet is the sum over
the calorimeter towers in the jet, where each  calorimeter tower is treated as a massless 4-vector, and the jet mass
is obtained from the resulting 4-vector.

We select events in a sample with \sampleLum\ integrated luminosity identified with an inclusive jet trigger requiring at least one jet with transverse energy ($\ET$) $>100$~GeV, with the trigger becoming fully efficient for jets with $\ET>140$~GeV.
Jet candidates are constructed with a Midpoint cone algorithm with cone radii of $R=0.4$\ and 0.7 and with the \antikt\ algorithm with a distance parameter $R=0.7$.
Primary collision vertices are reconstructed using charged particle information.
Events are required to have at least one high quality primary vertex with $|z_{vtx}|<60$~cm.  
Events are also required to be well-measured by requiring that they satisfy a  missing transverse energy significance requirement of $\mSigmet<10\ {\rm GeV}^{1/2}$, defined as 
\begin{eqnarray}
\mSigmet \equiv \frac{\met}{\sqrt{\sum \ET}},
\end{eqnarray}
where the sum is over all calorimeter towers.
We calculate for each jet the scalar sum of the $\pT$\ of the tracks associated with the jet cluster.
Each jet is required to either have more than $5\%$ of its energy registered in the electromagnetic calorimeter or to have
its summed track momentum be at least  5\%.  
This criterion eliminates jet candidates arising from instrumental backgrounds.
Furthermore, we restrict the jet candidates to have  $0.1 < |\eta_d| < 0.7$, where $\eta_d$\ is the jet pseudorapidity in the detector frame of reference, to ensure optimal calorimeter and charged particle tracking coverage.
We further require that the leading jet in the event have $\pT>400$~\GeVc.
We observe 2699 events.  

The jet 4-momentum is corrected to take into account calorimeter energy response, which is known to a precision of 3\%\  \cite{CDFJetNIM}\ for central calorimeter jets with  $\pT > 400$~\GeVc.  
We have determined the uncertainty on calibration of the jet mass measurement by comparing the momentum flux of charged particles into three concentric regions of the calorimeter around the jet centroid with the corresponding calorimeter response.

The number of interaction vertices ($N_{vtx}$) is a measure of the number of multiple interactions (MI), {\it i.e.} additional collisions in the same bunch crossing, and averages $\sim 3$\ in this sample.
We make a data-driven correction for MI effects on the jet substructure variables~\cite{Alon:2011xb}.
We select a subset of events with a back-to-back dijet topology.
We then define cones at right angles to the leading jet in azimuth of the same size as the jet cluster, and add the calorimeter towers in these cones to the jet 4-vector after rotation by $90^\circ$\ into the jet cone.  
The resulting average mass shift upward as a function of $\mjet{}$\  is taken as the correction downward due to MI and the energy flow from the underlying event (UE) of the hard collision.
We separately measure the UE correction by using only events with  $\Nvtx =1$.  
We correct the leading jet mass, $\mjet{1}$, for events with $\Nvtx>1$\ by the difference between the mass shift in multi-vertex events and the mass shift in single vertex events.
The correction has an approximate $1/\mjet{1}$\ behaviour and averages $\sim 4$~\GeVcc\ for a jet cone size of $R=0.7$.
The jet mass correction for a cone size of $R=0.4$\ is $\sim 0.5$~\GeVcc, consistent with the expected  $R^4$\ scaling \cite{Salam:2009jx}.
In the following, we focus on results for $R=0.7$\ Midpoint jets.


%
%

To model the high $\pT$\ processes, we used a \Pythia\ 6.216 calculation \cite{pythia}\ of QCD jet production generated with parton $\hat{p}_T > 300$~\GeVc, using the Tune A \cite{Field:2006} parameters for the underlying event and the CTEQ5L parton distribution functions (PDFs), followed
by a full detector simulation.
Based on a \Pythia\ calculation, we estimate $W$\ and $Z$\ boson production to contribute $\sim25$\ jets with masses between 60 and 100~\GeVcc,  which is less than 5\% of the number observed.
However, top quark pair production can contribute to the jet mass region $\mjet{1} > 100$~\GeVcc\ where the expected QCD jet rate is much lower.  
We employ an approximate
next-to-next-to-leading order (NNLO) calculation of the \ttbar\ differential cross section \cite{KidonakisVogt:2003}\ updated with the 
MSTW 2008 PDFs~\cite{Martin:2008MSTW}\ and a top quark mass of $m_{top}=173$~\GeVcc\ \cite{Kidonakis:2010PersCorr}.  This yields a cross section for top quark jets with 
$\pT>400$~\GeVc\ of $4.6$~fb.  
We used the \Pythia\ 6.216 generator to create a \ttbar\ MC sample and applied the same selection requirements used to define the event sample.
The estimated \ttbar\ contribution to the data sample, normalized to the NNLO cross section, is $13\pm4$~events.

Two-thirds of the \ttbar\ events with a leading high $\pT$\ jet would produce a recoil jet with a large jet mass ($\mjet{2}$) arising from the fully-hadronic decay of the recoil top quark.
The remaining \ttbar\  events  would have a recoil top quark that decays semileptonically, resulting in large $\MEt$\ and a recoil jet with lower $\pT$\ and $\mjet{2}$.
We reduce these backgrounds by rejecting events with $\mjet{2} > 100$~\GeVcc\ or by making a more
stringent \MEt\ requirement by rejecting events with $\mSigmet > 4\ {\rm GeV}^{1/2}$.
Approximately $25$\%\ (80\%) of the \ttbar\ (QCD) MC events survive these requirements.
We observe 30 jets with $\mjet{}>140$~\GeVcc\ and expect a \ttbar\ contribution of at most three jets.

In order to compare our results with QCD predictions, we correct the $\mjet{}$ distributions for effects of selection and resolution by an unfolding procedure, where we correct bin-by-bin 
the observed $\mjet{1}$\ distribution by the ratio of the QCD \Pythia\ MC $\mjet{1}$\ distribution without detector effects
and the same distribution after measurement and selection effects have been 
included.
This jet mass unfolding correction was derived for each jet algorithm separately, and the correction factors vary from 1.6 to 2.0 over the jet mass range $>70$~\GeVcc.
These corrections were verified through studies of the data and confirmed with MC calculations.


%
%

%
We summarize briefly our estimates of the systematic uncertainties that affect the substructure observables.
The overall jet mass scale at these energies is known to  2 (10)~\GeVcc\ for jet masses of 60 (120)~\GeVcc, based on the  jet energy scale uncertainty and the comparison of the calorimeter energy and track momentum measurements within the jet mentioned above.
We  assign an uncertainty on the MI correction of  2~\GeVcc, which is half of the average correction.
We assign a $\sim15$\%\ uncertainty on the jet  mass unfolding correction due to modeling of the jet hadronization, the uncertainty arising from the selection, and  MC statistical uncertainties.
The hadronization uncertainty is conservatively determined by comparing the change in the correction when  hadronization is turned off in the MC samples.
We estimate the PDF uncertainties on the \Pythia\ predictions  by reweighting the MC events using the $\pm 1 \sigma$\ variations in the  20  eigenvectors describing the uncertainties in the PDFs \cite{PDFUncertainties};
the uncertainties on the jet mass, angularity and planar flow distributions are 10\%\ or less in all cases.

We show in Fig.~\ref{Fig: Jet1MassR07Compared}\ a comparison of the unfolded $\mjet{1}$\ distribution for a cone size $R=0.7$\ with the analytic predictions for the jet function.
This comparison, made for jet masses above 70~\GeVcc, shows that the analytical prediction for quark jets describes approximately the shape of the distribution and fraction of jets but tends to over-estimate the rate for jet masses from 130 to 200~\GeVcc.  
The better agreement of the quark jet function with data compared with that of the gluon is consistent with the pQCD prediction that  $\sim80$\%\ of these jets arise from quarks \cite{Ellis:1994QCDCollider}.
Furthermore, the data and the \Pythia\ distributions are in reasonable agreement.
We also compare in the inset figure the distributions obtained for the Midpoint and \antikt\ algorithms.  
The \antikt\ jets have a very similar mass distribution to the Midpoint jets.
We find that $1.4\pm0.3$\%\ of the Midpoint jets with $\pT>400$~\GeVc\ have $\mjet{1}>140$~\GeVcc.
This is the first measurement of this rate, and allows us to constrain QCD predictions of this fraction, and provide the first measurement
of the rate of backgrounds in a massive jet sample from QCD production of high $\pT$\ light quarks and gluons.

\begin{figure}[t]
\center
 \leavevmode
 \resizebox{8.6cm}{5.7cm}
{\includegraphics[width=10cm]{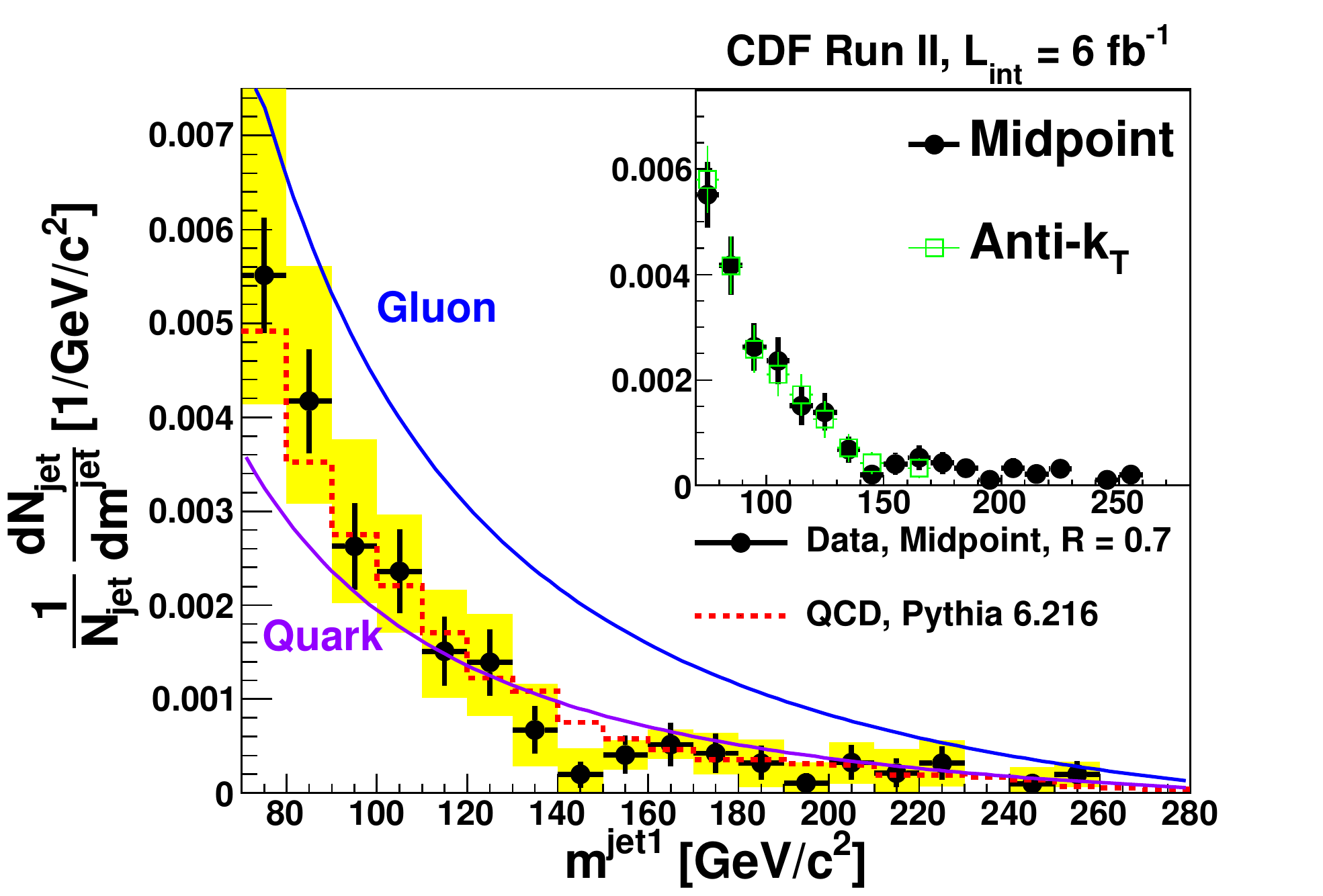} }
 \caption{  \label{Fig:  Jet1MassR07Compared}
The normalized jet mass distribution for Midpoint jets with $\pT > 400$~\GeVc\ and $|\eta|\in (0.1,0.7)$.  
The uncertainties shown are statistical (black lines) and systematic (yellow bars).
The theory predictions for the jet function for quarks and gluons are shown as solid curves  
and  have an estimated uncertainty of $\sim30$\%.
We also show the \Pythia\ MC prediction (red dashed line).
The inset compares Midpoint (full black circles) and \antikt\ (open green squares) jets.
}
\end{figure}

%
%
A key prediction of the NLO QCD calculation is that  the distribution of angularities  \cite{Berger:2003iw,Almeida:2008yp}\ of high mass jets 
has relatively sharp kinematical edges, with minimum and maximum values given by
\begin{equation}
\angularity^{\rm min} \sim (2/z)^{-3}, \, \angularity^{\rm  max} \sim zR^2/2^{3} ,
\end{equation}
 with $z\equiv \mjet{}/p_T\,.$
We show in  Fig.~\ref{Fig: AngularityC7Data}\ the angularity distribution  for the leading jet requiring that $\mjet{1} \in(90,120)$~\GeVcc.  The requirement of a relatively narrow $\mjet{1}$\ window allows us to compare the observed distribution with the shape and kinematic endpoints
predicted by pQCD. 
The \Pythia\ and QCD predictions are in good agreement with the data for Midpoint and \antikt\ jets, although the small size of the 
jet sample after applying the mass criterion limits the statistical precision of the comparison.
This further strengthens the interpretation that these massive jets arise from two-body configurations.
The small number of jets below $\angularity^{\rm min} $\ arise from resolution effects.
The PDF uncertainties on the \Pythia\ predictions are 10\%, and are shown in the figure.
The results for jets with  cone sizes of $R=0.4$\ are similar.  

\begin{figure}[t]
\center
 \leavevmode
 \resizebox{8.6cm}{5.7cm}
{\includegraphics[width=10cm]{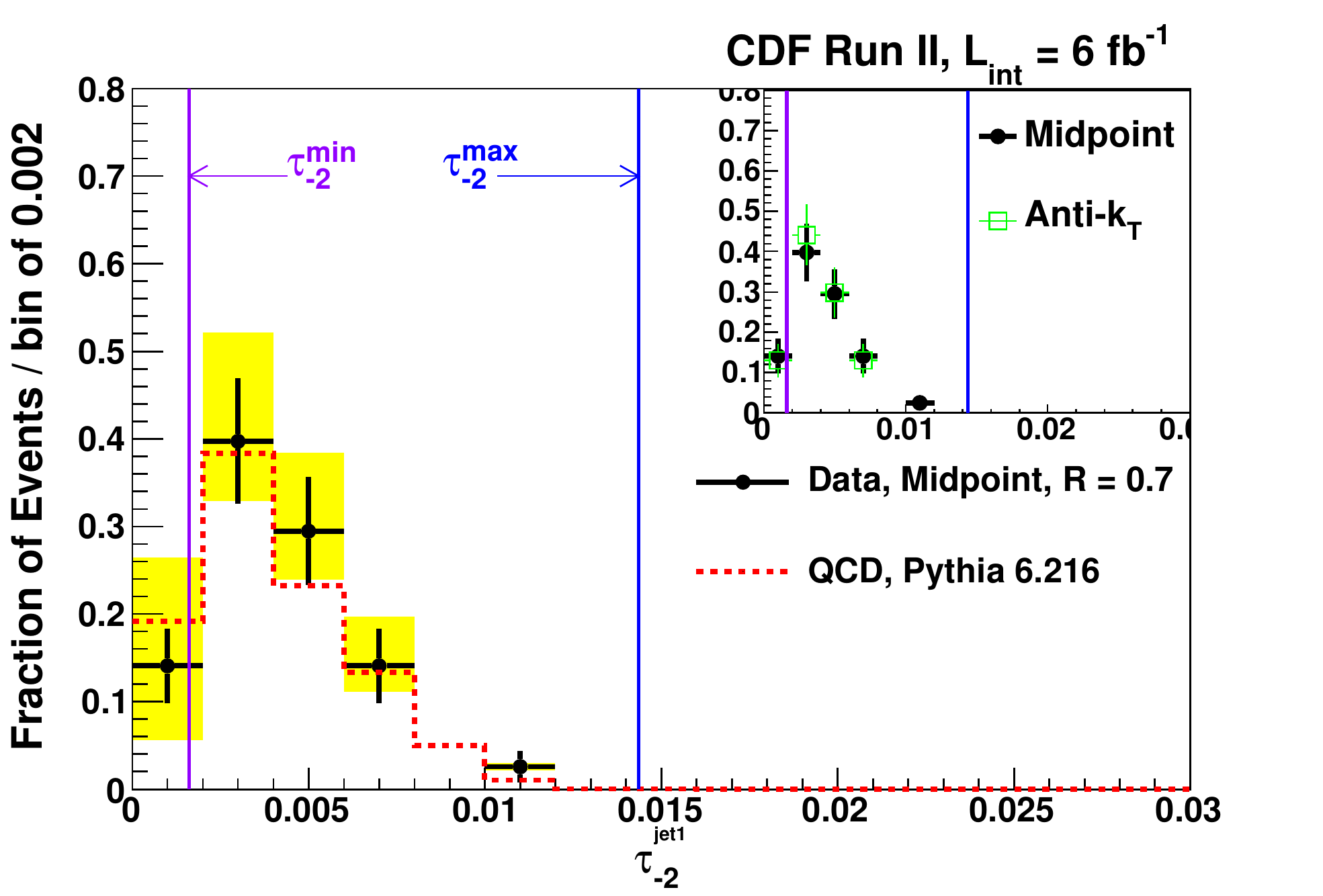} }
 \caption{  \label{Fig:  AngularityC7Data}
The angularity distribution for Midpoint jets with $\pT > 400$~\GeVc\ and $|\eta|\in (0.1,0.7)$.  
We have applied cuts to reject \ttbar\ events and required that  $\mjet{1} \in(90,120)$~\GeVcc.  
We also show the \Pythia\ calculation (red dashed line)\ and the pQCD kinematic endpoints.
The inset compares the distributions for Midpoint (full black circles) and \antikt\ (open green squares) jets.
}
\end{figure}

%
%
Figure~\ref{Fig: PlanarFlowC7Data}\ shows the planar flow distribution for jets where the jet mass is required to be in the range $130 - 210$~\GeVcc, relevant for jets arising from top quark decays.
Comparisons with the \Pythia\ predictions are also shown for both QCD multi-jet and \ttbar\ production.
Although the data are in good agreement with the predictions from QCD, the comparison is statistically limited because of the 
small number of observed jets in this jet mass range.
The PDF uncertainties on the \Pythia\ QCD predictions are 10\%.
The results for jets reconstructed with the Midpoint and \antikt\ algorithms are in good agreement with each other and are consistent with the general expectation based on MC calculations~\cite{Almeida:2008tp}.  
This study suggests that with higher statistics it will be possible to use the planar flow variable to discriminate high $\pT$\  QCD and top quark jets independent of jet mass.

\begin{figure}[t]
\center
 \leavevmode
 \resizebox{8.6cm}{5.7cm}
{\includegraphics[width=10cm]{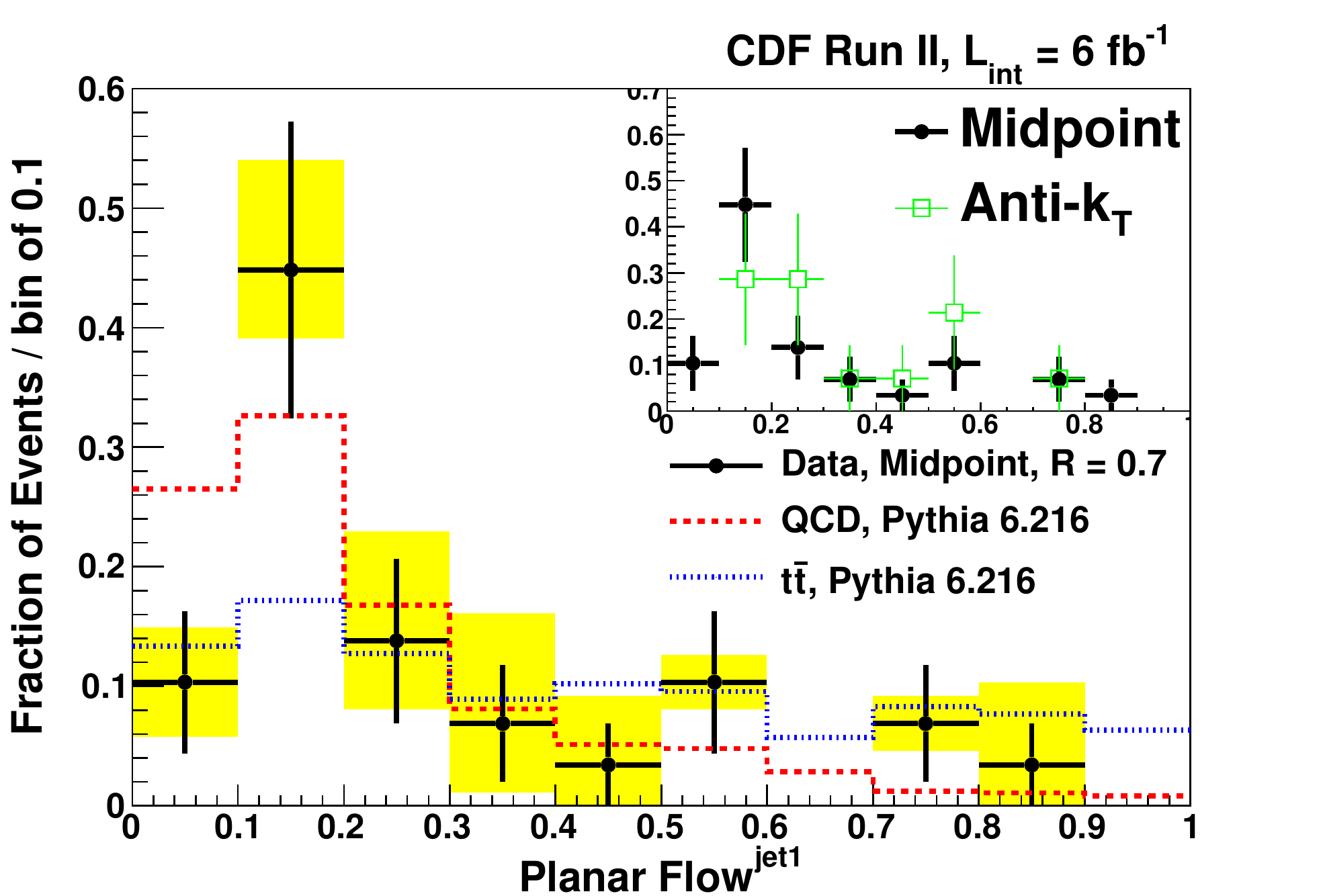} }
 \caption{  \label{Fig:  PlanarFlowC7Data}
The planar flow distributions for Midpoint jets with $\pT > 400$~\GeVc\ and $|\eta|\in (0.1,0.7)$\ 
after applying the top rejection cuts 
and requiring  $\mjet{1} \in(130,210)$~\GeVcc.
 We also show the \Pythia\ QCD  (red dashed line) and \ttbar\ (blue dotted line) jets, as well as the results from the two jet algorithms (inset).  All distributions have been separately normalized to unity.
 We expect only $\sim10$\%\ of the jets to arise from SM \ttbar\ production.
}
\end{figure}

In summary, we have measured for the first time the mass, angularity and planar flow distributions for jets with $\pT> 400$~\GeVc\ using Midpoint and \antikt\ jet algorithms.
We find good agreement between \Pythia\  Monte Carlo predictions, the NLO QCD jet function predictions, and the data for the jet mass distribution above 100~\GeVcc\ for Midpoint and \antikt\ jets.
The Midpoint and \antikt\ algorithms have very similar jet substructure distributions for high mass jets.
Our results show that the use of jet mass is an effective variable for separation of jets produced through QCD and through
\ttbar\ production, with a jet mass requirement of greater than 140~\GeVcc\ leaving only $1.4\pm0.3$\%\ of the QCD jets.
We have also shown that the high mass jets coming from light quark and gluon production are consistent with 
two-body final states from a study of the angularity variable, and that it may be possible to use the planar flow variable to further reject high mass QCD jets.  
These results provide the first experimental evidence that validates the MC calculations employing jet substructure to search for exotic heavy particles.



\begin{acknowledgments}

We acknowledge the contributions of I.~Sung and G.~Sterman for discussions involving non-perturbative effects in QCD jets, and to N.~Kidonakis for updated top quark differential cross section calculations. 

%
%
We thank the Fermilab staff and the technical staffs of the participating institutions for their vital contributions. This work was supported by the U.S. Department of Energy and National Science Foundation; the Italian Istituto Nazionale di Fisica Nucleare; the Ministry of Education, Culture, Sports, Science and Technology of Japan; the Natural Sciences and Engineering Research Council of Canada; the National Science Council of the Republic of China; the Swiss National Science Foundation; the A.P. Sloan Foundation; the Bundesministerium f\"ur Bildung und Forschung, Germany; the Korean World Class University Program, the National Research Foundation of Korea; the Science and Technology Facilities Council and the Royal Society, UK; the Institut National de Physique Nucleaire et Physique des Particules/CNRS; the Russian Foundation for Basic Research; the Ministerio de Ciencia e Innovaci\'{o}n, and Programa Consolider-Ingenio 2010, Spain; the Slovak R\&D Agency; the Academy of Finland; and the Australian Research Council (ARC). 

This work was supported in part by a grant from the Shrum Foundation, and by the Weizmann Institute of Science.

\end{acknowledgments}


\bibliography{QCDSubstructurePRL}

\end{document}